\documentstyle[aps,pre]{revtex}
\input epsf.sty
\begin{document}

\title{Anomalous scaling in two models of the passive scalar advection:\\
Effects of  anisotropy and compressibility}
\author{N.~V.~Antonov$^{1}$  and Juha~Honkonen$^{2}$\\
\small{$^{1}$ Department of Theoretical Physics, St Petersburg University,
Uljanovskaja 1, St Petersburg, Petrodvorez, 198904 Russia\\
\small$^2$ Theory~Division, Department~of~Physics, P.O.~Box~9
(Siltavuorenpenger~20C), FIN-00014 University~of~Helsinki,
Finland }}

\draft
\date{\today}
\maketitle

\begin{abstract}

The problem of the effects of compressibility and large-scale anisotropy
on anomalous scaling behavior is considered for two models describing
passive advection of scalar density and tracer fields. The advecting
velocity field is Gaussian, $\delta$-correlated in time, and scales
with a positive exponent $\varepsilon$.
Explicit inertial-range expressions for the scalar correlation
functions are obtained; they are represented by superpositions of
power laws with nonuniversal amplitudes and universal (dependent only
on $\varepsilon$ and $\alpha$, the compressibility parameter) anomalous exponents.
The complete set of anomalous exponents for the pair correlation
functions is found nonperturbatively, in any space dimension $d$,
using the zero-mode technique. For higher-order correlation functions,
the anomalous exponents are calculated to $O(\varepsilon^{2})$ using the
renormalization group. Like in the incompressible case, the exponents
exhibit a hierarchy related to
the degree of anisotropy: the leading contributions to the even
correlation functions are given by the exponents from the isotropic shell,
in agreement with the idea of restored small-scale isotropy.
As the degree of compressibility increases, the corrections become closer
to the leading terms.
The small-scale anisotropy reveals itself in the odd ratios of
correlation functions: the skewness factor is slowly decreasing going
down to small scales for the incompressible case, but becomes increasing
if $\alpha$ is large enough. The higher odd dimensionless ratios
(hyperskewness etc.) increase, thus signalling the persistent
small-scale anisotropy; this effect becomes more pronounced for
larger values of $\alpha$.
\end{abstract}

\section{Introduction}

Much attention has been paid recently to a simple model of the passive
scalar advection by a self-similar Gaussian white-in-time velocity field,
the so-called ``rapid-change model,'' introduced by Kraichnan \cite{K68};
see, e.g., Refs.~\cite{K94,Falk1,Falk2,GK,BGK,Pumir,RG,Gat} and
references therein.
Despite its simplicity, the model reproduces many of the anomalous
features of genuine turbulent heat or mass transport observed in
experiments.
On the other hand, it appears easier tractable theoretically: for the first
time, the anomalous exponents have been calculated on the basis of a
microscopic model and within regular expansions in formal small
parameters \cite{Falk1,Falk2,GK,BGK,Pumir,RG}. Therefore, the passive
advection by
the ``synthetic'' velocity with prescribed statistics, being of practical
importance in itself, may also be viewed as the starting point in studying
anomalous scaling in the turbulence on the whole.

In the original Kraichnan model, the velocity field is taken to be
Gaussian, isotropic, incompressible and decorrelated in time. More
realistic models should involve anisotropy and compressibility.
Recent studies have pointed out significant differences
between the compressible and incompressible cases
\cite{VM,El,Avell,tracer3,tracer,tracer2,AA98,A99}.
It is noteworthy that the
potential velocity field remains nontrivial in the
one-dimensional case, which is more accessible to numerical simulations
and allows interesting comparison between the numerical and analytical
results; see Ref.~\cite{VM}.

Another important question recently addressed is the effects of
large-scale anisotropy on inertial-range statistics of passively
advected scalar \cite{Sree,synth,Siggia,A98,ReG,CLMV99} and vector
\cite{LM99,ALM99,AHMM} fields and the
velocity itself \cite{Sadd,Arad}. These studies have shown that
the anisotropy present at large scales has a strong influence on
the small-scale statistical properties of the scalar, in
disagreement with what was expected on the basis of the cascade ideas
\cite{Sree,synth,Siggia}. On the other hand, the exponents describing
the inertial-range scaling exhibit universality and hierarchy related
to the degree of anisotropy, which gives some quantitative support to
Kolmogorov's hypothesis on the restored local isotropy of the
inertial-range turbulence
\cite{A99,A98,ReG,CLMV99,LM99,ALM99,AHMM,Sadd,Arad}.

In this paper we analyze the effects of the large-scale anisotropy
induced by a random source on a passive scalar by two methods:
First, we carry out the zero-mode calculation of the correlation
function of the passive scalar with an anisotropic source field
and in an
isotropic compressible velocity field decorrelated in time. Second,
for the same model, we have performed the two-loop renormalization-group
analysis of the
asymptotic behavior of the structure functions of the passive scalar
of arbitrary order.
This paper is organized as follows: In Sec. II, the zero-mode solution
for the correlation function both for passive density and passive tracer is
constructed. Two-loop renormalization-group analysis of the structure
functions of the passive scalar is carried out
in Sec. III with the use of the operator-product expansion.
Section IV is devoted to the discussion of the results.

\section{Zero-mode solution for passive density and tracer}

There are two types of diffusion-advection problems for the compressible
velocity field \cite{Landau}.
Passive advection of a density field $\theta(x)\equiv \theta(t,{\bf x})$
(say, the density of an impurity) is described by the equation
\begin{equation}
\partial _t\theta+ \partial_{i}(v_{i}\theta)
=\nu _0\partial^{2} \theta+f,
\label{density1}
\end{equation}
while the advection of a ``tracer'' (say, temperature, specific entropy, or
{\it concentration} of the impurity particles) is described by
\begin{equation}
\partial _t\theta+ (v_{i}\partial_{i})\theta
=\nu _0\partial^{2} \theta+f.
\label{tracer1}
\end{equation}
Here $\partial _t \equiv \partial /\partial t$,
$\partial _i \equiv \partial /\partial x_{i}$, $\nu _0$
is the molecular diffusivity coefficient, $\partial^{2}$
is the Laplace operator, ${\bf v}(x)$ is the velocity field,
and $f\equiv f(x)$ is an artificial Gaussian scalar noise with zero
mean and the covariance
\begin{equation}
\langle f(x)f(x') \rangle = \delta(t-t')\, C({\bf r}), \qquad
{\bf r} \equiv {\bf x} - {\bf x}' ,
\label{noise}
\end{equation}
where $C({\bf r})$ varies noticeably on $r\equiv |{\bf r} |\sim L$,
the integral turbulence scale. In the presence of a preferred
direction specified by a unit vector ${\bf n} $, the function $C({\bf r})$
can be written in the form
\begin{equation}
C({\bf r})= \sum_{l=0}^{\infty}  C_{l} (mr) \, P_{l} (z) ,
\quad z\equiv \frac{{\bf r} \cdot{\bf n} }{r} ,
\label{noise2}
\end{equation}
where $m\equiv 1/L$, $C_{l} (mr)$ are coefficient functions
such that $C({\bf r})$ becomes constant
at $mr=0$ and decays rapidly for $mr\to\infty$,
$z$ is the cosine of the angle between ${\bf n} $ and ${\bf r} $,
and $P_{l} (z)$ are ($d$-dimensional)
Legendre polynomials satisfying the equations
\begin{equation}
(1-z^{2}) P_{l}'' (z) + z\,(1-d) P_{l}'(z) +l (l+d-1) P_{l}(z) =0.
\label{Legendre}
\end{equation}

The anisotropy makes it possible to introduce also a mixed correlator
$\langle{\bf v}f\rangle\propto{\bf n}  \delta(t-t')\, C'({\bf r} )$ with some
function
$C'$ similar to $C$ in Eq. (\ref{noise2}). This violates the
evenness in ${\bf n} $ and gives rise to nonvanishing odd correlation functions
of $\theta$, but leads to no serious alterations in the analysis.
We shall discuss this case later on, and for the time being we assume
$\langle{\bf v}f\rangle=0$.

In the real problem, the field ${\bf v}(x)$ satisfies the Navier--Stokes
equation. In the simplified model considered in
\cite{K68,K94,Falk1,Falk2,GK,BGK,Pumir} it obeys a
Gaussian distribution with zero mean and the covariance
\begin{equation}
\bigl\langle v_{i}(x) v_{j}(x')\bigr\rangle = \delta(t-t')\,
K_{ij}({\bf r} )
\label{temporal}
\end{equation}
with
\begin{equation}
K_{ij} ({\bf r} ) =
\int \frac {d{\bf k}}{(2\pi)^d}\,
\frac{D_{0}\,P_{ij}({\bf k})+D_{0}'\, Q_{ij}({\bf k})}
{(k^{2}+m^{2})^{d/2+\varepsilon/2}} \,
\exp [{\rm i}({\bf k}\cdot{\bf r})] ,
\label{spatial}
\end{equation}
where $P_{ij}({\bf k}) = \delta _{ij} - k_i k_j / k^2$
and $Q_{ij}({\bf k}) = k_i k_j / k^2$ are the
transverse and longitudinal projectors, respectively,
$k\equiv |{\bf k}|$, $D_{0}$ and $D_{0}'$ are positive amplitude
factors, and $d$ is the dimensionality of the coordinate space.
For $D_{0}'=0$ (the incompressible case) the models (\ref{density1}) and
(\ref{tracer1}) coincide. For $0<\varepsilon<2$, the so-called eddy diffusivity
\begin{equation}
S_{ij}({\bf r} ) \equiv K_{ij} ({\bf 0}) - K_{ij} ({\bf r} )
\label{difference}
\end{equation}
has a finite limit for $m\to0$:
\begin{equation}
S_{ij}({\bf r} )= D r^{\varepsilon}\left [\left (d+\varepsilon -1+\alpha \right)
\delta_{ij} - \varepsilon (\alpha-1) \frac{r_{i}r_{j}}{r^2} \right],
\label{eddydiff}
\end{equation}
with
\begin{equation}
 D = \frac{- D_{0}\, \Gamma(-\varepsilon /2)} {(4\pi)^{d/2}2^{\varepsilon }(d+\varepsilon )
\Gamma(d/2+\varepsilon /2)}, \quad
\alpha \equiv  D_{0}' /D_{0} ,
\label{amplitudes}
\end{equation}
where $\Gamma$ is the Euler gamma function (note that $D$ and
$\alpha$ both are positive). In the renormalization group (RG)
approach, the exponent $\varepsilon $ plays
the same role as the parameter $\varepsilon=4-d$ does in the RG theory
of critical behavior; see Refs.~\cite{RG,AA98}. The relation
$D_{0}/\nu_0 \equiv  \Lambda^{\varepsilon }$ defines the
characteristic ultraviolet wave-number scale $\Lambda$.

The issue of interest is the behavior of various correlation functions
in the inertial range specified by the inequalities $\Lambda r \gg 1$,
$mr\ll 1$.
In the models (\ref{density1})--(\ref{eddydiff}), odd multipoint
correlation
functions of the scalar field vanish, while the even equal-time
functions satisfy linear partial differential equations; see, e.g.,
\cite{K68,K94,Falk1,Falk2,GK,BGK,Pumir}. The equation for
the equal-time pair correlation function
$D({\bf r} )\equiv\langle \theta(t,{\bf x})\theta(t,{\bf x}') \rangle$,
is easily derived from the Schwinger--Dyson equations
(see Refs.~\cite{AA98,ALM99}) and has the form (here and below in equal-time
functions, we omit common to all the quantities time arguments):
\begin{equation}
2\nu_0 \partial^{2} D({\bf r} )+ \left[S_{ij}(r)\partial_{i}\partial_{j}
\right] D({\bf r} ) = C({\bf r} )
\label{master-t}
\end{equation}
for the model (\ref{tracer1}) and
\begin{equation}
2\nu_0 \partial^{2} D({\bf r} )+ \partial_{i}\partial_{j}
 \left[S_{ij}(r)D({\bf r} )\right] = C({\bf r} )
\label{master-d}
\end{equation}
for the model (\ref{density1}), with $C({\bf r} )$ from (\ref{noise2})
and $S_{ij}(r)$ from (\ref{eddydiff}).
In the presence of the preferred direction, ${\bf n} $, the correlation function can be
decomposed in the Legendre polynomials,
\begin{equation}
D({\bf r} ) = \sum_{l=0}^{\infty}  D_{l} (r) \, P_{l} (z),
\label{pair}
\end{equation}
where the coefficient functions are sought in the powerlike form
\begin{equation}
 D_{l} (r) \simeq  D_{l}\, r^{\zeta_{l}} .
\label{pair-power}
\end{equation}
Owing to the evenness in ${\bf n} $, only even polynomials contribute to
(\ref{pair}).

It is well-known (see, e.g., Refs.~\cite{Falk1,Falk2,GK,BGK})
that the nontrivial inertial-range exponents are determined by the
zero modes, i.e., solutions of Eqs. (\ref{master-t}), (\ref{master-d})
neglecting both the forcing [$C({\bf r} )=0$] and the dissipation ($\nu_0=0$).
The homogeneous equations are $SO(d)$ covariant, and the equations for
the coefficient functions in (\ref{pair}) foliate.
Substituting the representations (\ref{pair}), (\ref{pair-power}) into
Eqs. (\ref{master-t}), (\ref{master-d}) and using the relations
$\partial_{i} S_{ij} (r) = \alpha\varepsilon (d+\varepsilon )D r_{j} r^{-2+\varepsilon }$,
$\partial_{i}\partial_{j} S_{ij} (r) = \alpha\varepsilon (d+\varepsilon )(d-2+\varepsilon )D
r^{-2+\varepsilon }$ then gives quadratic equations for the exponents
$\zeta_{l}$ in Eq. (\ref{pair-power}), namely,
\begin{equation}
\zeta_{l}(\zeta_{l}+d-2) - l(l+d-2) +
\frac{\zeta_{l}(\zeta_{l}-1)\varepsilon (\alpha-1)}
{(d-1+\alpha+\varepsilon )} =0
\label{kvadrat-t}
\end{equation}
for the tracer, which has two solutions, $\zeta_{l}=2-d-l + O(\varepsilon )$ and
\begin{equation}
\zeta_{l} =l -\varepsilon \frac{l(l-1)(\alpha-1)}
{(d+2l-2)(d-1+\alpha)} + O(\varepsilon ^{2}),
\label{kvadrat-te}
\end{equation}
and
\begin{equation}
\zeta_{l}(\zeta_{l}+d-2) - l(l+d-2) + \varepsilon \,
\frac{\zeta_{l}(\zeta_{l}-1)\,(\alpha-1) +
\alpha\, (d+\varepsilon )(2\zeta_{l} +d-2+\varepsilon )}
{(d-1+\alpha+\varepsilon )} =0
\label{kvadrat-d}
\end{equation}
for the density, with two solutions, $\zeta_{l}= -d-l + O(\varepsilon )$ and
\begin{equation}
\zeta_{l} =l -\varepsilon \frac{l(l-1)(\alpha-1)+\alpha d(d+2l-2) }
{(d+2l-2)(d-1+\alpha)} + O(\varepsilon ^{2}).
\label{kvadrat-de}
\end{equation}

The standard arguments \cite{Falk1,Falk2,GK,BGK}
show that only the second
solution, $\zeta_{l} =l+O(\varepsilon )$, is ``admissible.'' It has the form
\begin{eqnarray}
\label{root-te}
\zeta_{l}=
[2(-1+d+\alpha+\alpha\varepsilon)]^{-1}
\left\{-2+3\,d-{d^2} + 2\,\alpha  - d\,\alpha  +
\varepsilon- d\,\varepsilon+\alpha\,\varepsilon\right.\nonumber\\
+
[{{( 2 - 3\,d + {d^2} - 2\,\alpha  +
              d\,\alpha  - \varepsilon  + d\,\varepsilon  -
              \alpha \,\varepsilon  ) }^2}\\
              \left. -
         4\,l\,( -2 + 3\,d - {d^2} + l - d\,l +
            2\,\alpha  - d\,\alpha  - l\,\alpha  +
            2\,\varepsilon  - d\,\varepsilon  - l\,\varepsilon
             ) \,( -1 + d + \alpha  +
            \alpha \,\varepsilon  ) ]^{1/2}\right\}\nonumber
\end{eqnarray}
for the tracer and
\begin{eqnarray}
\label{root-de}
\zeta_{l}=[2\,\left( -1 + d + \alpha  +
       \alpha \,\varepsilon  \right) ]^{-1}\left\{-2 + 3\,d - {d^2} + 2\,\alpha  - d\,\alpha  +
\varepsilon  -
     d\,\varepsilon  + \alpha \,\varepsilon  - 2\,d\,\alpha \,\varepsilon  -
     2\,\alpha \,{{\varepsilon }^2}\right.\nonumber\\ +
     \left[{{\left( 2 - 3\,d + {d^2} - 2\,\alpha  + d\,\alpha  -
              \varepsilon  + d\,\varepsilon  - \alpha \,\varepsilon  +
              2\,d\,\alpha \,\varepsilon  + 2\,\alpha \,{{\varepsilon }^2}
               \right) }^2} -
         4\,\left( -1 + d + \alpha  + \alpha \,\varepsilon  \right)\right.\nonumber\\
         \times
          \left( -2\,l + 3\,d\,l - {d^2}\,l + {l^2} - d\,{l^2} +
            2\,l\,\alpha  - d\,l\,\alpha  - {l^2}\,\alpha  +
            2\,l\,\varepsilon  - d\,l\,\varepsilon  - {l^2}\,\varepsilon  -
            2\,d\,\alpha \,\varepsilon\right.\\\left.\left.  + {d^2}\,\alpha \,\varepsilon  -
            2\,\alpha \,{{\varepsilon }^2} +
            2\,d\,\alpha \,{{\varepsilon }^2} + \alpha \,{{\varepsilon }^3}
             \right) ]^{1/2}\right\}\nonumber
\end{eqnarray}
for the density. For $\alpha=0$, these solutions coincide with each other and with the
exponents obtained earlier in Refs.~\cite{Falk1,Gat} for the
incompressible case.

The exponents (\ref{root-te}), (\ref{root-de}) exhibit a
hierarchy related to the degree of anisotropy:
\begin{equation}
\zeta_{l}>\zeta_{l'} \quad {\rm if} \quad l>l' ,
\label{hier-2}
\end{equation}
i.e., the less is the index $l$, the less is the exponent and,
consequently, the more important is the contribution to the inertial-range
behavior. The leading term is given by the exponent $\zeta_{0}$ from the
``isotropic shell.'' This behavior is illustrated by Figs.~1 and~2.

Although the hierarchy holds for all values of $\alpha$
($\partial \zeta_{l} / \partial l>0$), the corrections
become closer to leading terms as $\alpha$ increases:
$\partial^{2} \zeta_{l} / \partial l \partial\alpha<0$.
This behavior is illustrated by Fig.~3.

Since the equation (\ref{tracer1}) is invariant with respect to the shift
$\theta\to\theta+{\rm const} $, the relevant quantities for the tracer are the
so-called structure functions,
\begin{equation}
S_{n}({\bf r} )= \langle [\theta(t,{\bf x})-\theta(t,{\bf x}')]^{n}\rangle,
\qquad    {\bf r} = {\bf x}-{\bf x}'  ,
\label{struc}
\end{equation}
with the Legendre decomposition
\begin{equation}
S_{n}({\bf r} )= \sum_{l=0}^{\infty} S_{nl} \, r^{\zeta_{nl}} \, P_{l} (z)
\label{struc-Le}
\end{equation}
with some numerical coefficients $S_{nl}$. The comparison with Eqs.
(\ref{pair}), (\ref{pair-power}) gives $\zeta_{2l}=\zeta_{l}$ with
$\zeta_{l}$ from (\ref{root-te}) for all $l>0$. For $l=0$, the constant
term with $\zeta_{0}=0$ drops out from the difference in (\ref{struc}),
and the behavior of the isotropic shell is determined by the
subleading exponent $(2-\varepsilon )$; see, e.g., Ref.~\cite{A99}. Note
that the hierarchy relations (\ref{hier-2}) remain valid also for
$S_{2}=[D({\bf 0})-D({\bf r} )]/2$.

For the density case, the exponent
$$ \zeta_{0}=\frac{-\varepsilon (d+\varepsilon )\alpha}{(d-1)+\alpha(1+\varepsilon )}<0 $$
(the square root in Eq. (\ref{root-de}) is taken explicitly)
gives the leading contribution both for the pair correlation function
$D({\bf r} )$ and the structure function $S_{2}$ in  (\ref{struc}), in
agreement
with the exact solution of \cite{AA98} (for $d=1$, see Ref.~\cite{VM}).
Note that for this case, the anomalous scaling emerges already for the
pair correlation function, like in the model
of passively advected magnetic field studied in Ref.~\cite{V96}.

\section{Two-loop renormalization-group analysis of structure functions}

The higher-order structure functions can be studied using the field
theoretic renormalization group and operator product expansion.
The detailed exposition of these techniques and practical calculations
can be found in Refs.~\cite{RG,AA98,A99,A98,ALM99};
below we confine ourselves to only the necessary information.

The field theoretic models corresponding to the stochastic
equations (\ref{density1}) and
(\ref{tracer1})  are multiplicatively renormalizable; the
corresponding RG equations have infrared stable fixed points. In
particular,
this leads to the following representations for the structure functions
in the
model (\ref{tracer1}) in the inertial range ($\Lambda r\gg1$, $mr\ll 1$):
\begin{equation}
S_{n} = D_{0}^{-n/2} r^{n(1-\varepsilon /2)} \sum_{a} C_{a} (r,z) (mr)^{\Delta_{a}}.
\label{RGR}
\end{equation}
Here $C_{a} (r,z)$ are coefficients analytical in $m$ and finite for
$m\to0$, and $\Delta_{a}$ are the critical dimensions of the composite
operators entering in the operator product expansion.

The leading zero-mode contribution in the $l$-th shell for $S_{n}$ is
determined by the critical dimension $\Delta_{nl}$ of the
irreducible traceless $l$-th rank tensor operator built of $n$ fields
$\theta$ and minimal possible number of derivatives. For $l\le n$ such
an operator has the form
\begin{equation}
\partial_{i_{1}}\theta\cdots\partial_{i_{l}}\theta\,
(\partial_{i}\theta\partial_{i}\theta)^{p}+\cdots, \quad n=l+2p.
\label{Fnp}
\end{equation}
Here the dots stand for the appropriate subtractions involving the
Kronecker delta symbols, which ensure that the resulting expressions are
traceless with respect to contraction of any given pair of indices, for
example, $\partial_{i}\theta\partial_{j}\theta - \delta_{ij}
\partial_{k}\theta\partial_{k}\theta /d$,
$ \partial_{i}\theta\partial_{j}\theta \partial_{k}\theta -
(\delta _{ij}\partial_{k}\theta  + \delta _{ik}\partial_{j}\theta +
\delta _{jk}\partial_{i}\theta) /(d+2)$ and so on.

The exponents $\zeta_{nl}=n(1-\varepsilon /2)+\Delta_{nl}$ are calculated
in the form of series in $\varepsilon $, where
$\zeta_{nl}=n+\sum _{k=1}^{\infty} \zeta_{nl}^{(k)}\varepsilon ^{k}$.
In the first order in $\varepsilon $:
\begin{equation}
\zeta_{nl}^{(1)} = - \frac{1}{(d+2)}\, \left\{ \frac{(n-l)(d+n+l)}{2} +
\frac{l(l-1)(\alpha-1)+\alpha(n-l)(n+l-2)}{(d-1+\alpha)}
 \right\}.
\label{Dnp2}
\end{equation}

For the incompressible model, the exponent from the isotropic shell
($l=0$) was obtained in Refs.~\cite{Falk1,Falk2} to the order $O(1/d)$ and
in Refs.~\cite{GK,BGK} to the order $O(\varepsilon)$; the results for
$n=3$ are given in Ref.~\cite{Pumir}. The general case can be found in
Ref.~\cite{A98}; see also \cite{ReG,KJW}. For general $\alpha>0$,
the exponent $\zeta_{n0}$ was found in Ref.~\cite{tracer}
(see also \cite{tracer2}; in the notation of those papers,
$\wp=\alpha/(d-1+\alpha)$). The result for general $n$, $l$ is given
in Ref.~\cite{A98} (for more details, see \cite{A99}).

We have performed the two-loop calculation of the exponents
$\zeta_{nl}$ and obtained:
\begin{eqnarray}
\label{eps-kvadrat}
\zeta_{nl}^{(2)}&=&
{[-d(d+1)+(d^2-2d-4)\alpha+(3d+4)\alpha^2](d\kappa_1-\kappa_2)\over
d(d+2)^3(d-1+\alpha)^2}\nonumber\\
&+&{(n-2)\over
4d(d+2)^3(d+4)(d-1+\alpha)^2}
\Bigl(8d[-(d^2+5d+10)-(d-2)(d+4)\alpha+(2d^2+7d+2)\alpha^2]\kappa_1\nonumber\\
&+&4[(d+1)(3d^3+17d^2+20d-24)-(d+4)(d^3+7d^2-2d-4)\alpha+(d+1)(5d^2+8d-24)\alpha^2]\kappa_2\\
&+&3(d+2)h(d)
\Bigl\{4d[3+(d-4)\alpha-(d-1)\alpha^2]\kappa_1
+[-3(d+1)(d^2+5d-4)\nonumber\\
&+&2(d^3+10d^2-d-4)\alpha-3(2d^2-3d-4)\alpha^2]\kappa_2\Bigr\}\Bigr)\nonumber
\end{eqnarray}
where we have written $\kappa_{1}=n(n-1)$, $\kappa_{2}=
(n-l)(n+l-2+d)$, $h(d) = F(1,1,d/2+1;1/4)$ and $F(a,b;c;z)$
is the hypergeometric function; see, e.g., Ref.~\cite{Gamma}.
For integer space dimension $d$ one has
$h(1)=2\pi/(3\sqrt 3)$ and $h(2)=4\ln(4/3)$,
and the others can be obtained from the recursion relation
\begin{equation}
3 h(d)+{d\over d+2}h(d+2)=4,
\label{recursion}
\end{equation}
valid for all $d$.

For the incompressible case, $\alpha=0$, Eq. (\ref{eps-kvadrat}) becomes:
\begin{eqnarray}
\label{kvadrat-tr}
\zeta_{nl}^{(2)}&=&
-{(d+1)(d\kappa_1-\kappa_2)\over
(d+2)^3(d-1)^2}+{(n-2)\over
4d(d+2)^3(d+4)(d-1)^2}
\Bigl\{-8d(d^2+5d+10)\kappa_1\nonumber\\
&+&4[(d+1)(3d^3+17d^2+20d-24)]\kappa_2
+9(d+2)h(d)
[4d\kappa_1
-(d+1)(d^2+5d-4)\kappa_2]\Bigr\}
\end{eqnarray}

The results for $l=0$ and 2 were obtained earlier in Ref.~\cite{RG}.
In that paper, they were expressed in terms of the functions
$h_{1} (d) = h(d+2)$ and $h_{2} (d) = h(d+4)$, which can be reduced
to the form (\ref{kvadrat-tr}) using the relation (\ref{recursion}).

For $d=1$, only the exponents $\zeta_{n0}$ and $\zeta_{n1}$ make sense
(traceless operators of the rank $l\ge2$ vanish identically). They are
independent of $\alpha$ (the velocity is purely potential) and have
the form  ($l=0$ if $n$ is even and $l=1$ if $n$ is odd)
\begin{equation}
\zeta_{n}=n+n{\varepsilon \over 2} -{n^{2}\varepsilon \over 2}   +
{n(n-1)(n-2)\varepsilon ^{2}\pi\over 6\sqrt{3}} +O(\varepsilon ^{3}).
\label{d1}
\end{equation}
The expression (\ref{d1}) is in agreement with the $O(\varepsilon )$ result of
Ref. \cite{VM} and the $O(\varepsilon ^{2})$ result of Ref.~\cite{AA98}
(in the
latter, the density case was studied, but in one dimension these models
can be related by the replacement $\theta \to\partial\theta$).

For $l>n$, the leading operators contain more derivatives than fields,
the corresponding exponents behave as $\zeta_{nl} = l-n+O(\varepsilon )$ and
the corresponding terms in representation (\ref{struc-Le}) rapidly decrease
for $mr\to0$.

Now let us return to the density model (\ref{density1}). It is not
invariant
with respect to the shift $\theta\to\theta+{\rm const} $, the operators
$\theta^{n}$ have nontrivial critical dimensions,
\begin{eqnarray}
\Delta_{n}= n\left(-1+{\varepsilon \over 2}\right)-\frac{\alpha n(n-1)d\varepsilon }{2(d-1+\alpha)}+
\frac{\alpha(\alpha-1) n(n-1)(d-1)\varepsilon ^{2}}{2(d-1+\alpha)^{2}}
+\frac{\alpha^{2} n(n-1)(n-2)d\, h(d)\varepsilon ^{2}}{4(d-1+\alpha)^{2}}+
O(\varepsilon ^{3}),
\label{Dn}
\end{eqnarray}
and different terms in the structure functions have
different scalings, see Ref.~\cite{AA98}. The relevant quantities are then
the  equal-time pair-correlation functions of the powers of $\theta$,
which have the form
\begin{equation}
\langle\theta^{n}({\bf x})\theta^{p}({\bf x'})\rangle\propto
\nu_0^{-(n+p)/2}\,\Lambda^{-(n+p)}\,
(\Lambda r)^{-\Delta_{n}-\Delta_{p}} \, F_{np}({\bf r})
(Mr)^{\Delta_{n+p}}.
\label{as1}
\end{equation}
The leading terms in the Legendre decompositions for the scaling functions,
\begin{equation}
F_{np}({\bf r}) = \sum_{l=0}^{\infty}  C_{l}^{(np)} (mr) \, P_{l} (z) ,
\label{as10}
\end{equation}
are given by the contribution from the scalar operators $\theta^{n+p}$
without derivatives, $C_{0}^{(np)} \propto (mr)^{\Delta_{n+p}}$.

The operators that determine corrections ($l>0$) to the leading term
with $l=0$ necessarily contain $l$ derivatives and, in contrast to the
tracer case, the correction exponents differ from the leading ones by
a few unities, $l+O(\varepsilon )$.
In particular, the leading $l=2$ correction is related to the operator
\[ \Bigl(\partial_{i}\theta\partial_{j}\theta-\delta_{ij}
(\partial_{k}\theta\partial_{k}\theta)/d \Bigr) \,
\theta^{n+p-2}, \]
whose dimension equals to
\[ 2+(n+p)(-1+\varepsilon /2) - \frac{\varepsilon }{(d-1+\alpha)} \left\{
\frac{(n+p)(n+p-1)\alpha d}{2}+\frac{2(\alpha-1)}{(d+2)}\right\}+
O(\varepsilon ^{2}). \]

Note also that the exponents $\zeta_{l}$ in Eqs. (\ref{root-te}),
(\ref{root-de}) are related to the composite operators with two fields
$\theta$ and $l$ free indices, which (up to total derivatives and
subtractions with the delta symbols) reduce to the form
$\theta \partial_{i_{1}}\cdots \partial_{i_{1}} \theta$; the one-loop
calculation confirms the $O(\varepsilon )$ results (\ref{kvadrat-te}),
(\ref{kvadrat-de}).

Since the leading terms of the even functions (\ref{struc-Le}) are
determined by the exponents of the isotropic shell (i.e., those related
to scalar composite operators), the inertial-range behavior of the former
is the same as in the isotropic model. This gives quantitative
support both to Kolmogorov's hypothesis
on the restored local isotropy of the inertial-range turbulence
and to the universality of anomalous exponents with respect to
the way in which the turbulence is excited.

On the contrary, the small-scale anisotropy reveals itself in odd
correlation
functions  (which are nonzero in the presence of a mixed correlator
$\langle{\bf v}f\rangle$ or the constant mean gradient of the scalar
field).
It follows from the above analysis that the dimensionless ratios
${\cal R}_{n}\equiv S_{2n+1} /S_{2}^{(2n+1)/2}$ for the tracer case
in the inertial range have the form
\begin{equation}
{\cal R}_{n} \propto (mr) ^{\Delta_{2n+1,1}} ,
\label{ratios}
\end{equation}
where $\Delta_{2n+1,1}=\zeta_{2n+1,1}- (2n+1)(2-\varepsilon )$ is the
critical dimension of the {\it vector} composite operator (\ref{Fnp})
built of $2n+1$ scalar gradients. From Eq. (\ref{Dnp2})
we find, to the first order in $\varepsilon $,
\begin{equation}
\Delta_{2n+1,1}= \varepsilon \,
\bigl[ (d-1+\alpha)(d+2-4n^{2})-8\alpha\, n^{2}\bigr] /2(d+2)(d-1+\alpha).
\label{ratios-2}
\end{equation}

For small $\alpha$, the skewness factor ${\cal R}_{1}$ decreases for
$mr\to0$, but slower than expected on the basis of cascade ideas (the
latter suggest that the odd ratios should vanish for small $mr$, where the
turbulence is expected to become isotropic), while the higher-order ratios
increase, thus signalling the persistence of small-scale anisotropy. When
$\alpha$ increases, ${\cal R}_{1}$ also becomes divergent for
$mr\to0$ provided $\alpha$ is large enough [namely,
$\alpha>(d-1)(d+2)/(10-d)+O(\varepsilon )$],
while the higher-order ratios diverge even faster.

It was argued in Refs. \cite{tracer3,tracer,tracer2} that
the anomalous scaling regime in the models at hand breaks down if
$\varepsilon $ and $\alpha$ are both large enough
[$\wp\equiv \alpha/(d-1+\alpha) > d/\xi^2$] and the inverse energy cascade
with no anomalous scaling takes place. This effect obviously cannot be
detected within the $\varepsilon $ expansion. It is noteworthy that the exact
nonperturbative exponents (\ref{root-te}), (\ref{root-de}) show no hint
of anomaly at the threshold, $\wp= d/\xi^2$, in contrast to the exact
exponents for the magnetic case which become complex when the anomalous
scaling regime breaks down; see Refs.~\cite{V96,LM99}.

\section{Conclusion}

To conclude with, we have studied
the effects of compressibility and large-scale anisotropy
on the anomalous scaling behavior in two models,
which describe
passive advection of scalar tracer and density fields. The advecting
velocity field is Gaussian, $\delta$-correlated in time, and its spatial correlations scale with
a positive exponent $\varepsilon $.
Explicit inertial-range expressions for the scalar correlation
functions have been obtained; they are represented by superpositions of
power laws with nonuniversal amplitudes and universal (dependent only
on $\varepsilon $ and $\alpha$, the compressibility parameter) anomalous exponents.
The complete set of anomalous exponents for the pair correlation
functions has been found nonperturbatively, in any space dimension $d$,
using the zero-mode approach. For higher-order correlation functions,
the anomalous exponents have been calculated to $O(\varepsilon ^{2})$ using the
RG techniques. Like in the incompressible case, the exponents
exhibit a hierarchy related to the degree of anisotropy; the leading
contributions to the even correlation functions are given by the exponents
from the isotropic shell, in agreement with the idea of restored
small-scale isotropy.

This picture seems rather general, being compatible with that established
recently for the Navier--Stokes turbulence \cite{Arad}, passive scalar
advection by the two-dimensional Navier--Stokes field \cite{CLMV99}, and
passive advection of a scalar \cite{A98} and vector \cite{LM99,ALM99}
fields by the white-in-time incompressible synthetic velocity field.

As the degree of compressibility increases, the corrections become closer
to the leading terms; cf. Ref.~\cite{AHMM} for the passive advection of a
magnetic field.

On the contrary, the small-scale anisotropy reveals itself in the odd
ratios of correlation functions: the skewness factor is slowly decreasing
down to small scales for the incompressible case \cite{Pumir}, but
becomes increasing if $\alpha$ is large enough. The higher-order odd
dimensionless ratios (hyperskewness etc.) increase, thus signalling
the persistent small-scale anisotropy, cf. Refs.~\cite{A99,A98,CLMV99}.
This effect becomes even more pronounced for larger values of $\alpha$;
cf. Ref.~\cite{AHMM} for the magnetic case.

\acknowledgements
It is a pleasure to thank L.~Ts.~Adzhemyan, A.~Mazzino,
P.~Muratore-Ginanneschi and S.~V.~Novikov for useful discussions.
N.\,V.\,A. acknowledges the Department of Physics of the University of
Helsinki for hospitality. The work was supported by
the Academy of Finland (Project No. 46439),
Grant Center for Natural Sciences (Grant No. 97-0-14.1-30) and Russian
Foundation for Fundamental Research (Grant No. 99-02-16783).

\begin{figure}
\epsffile{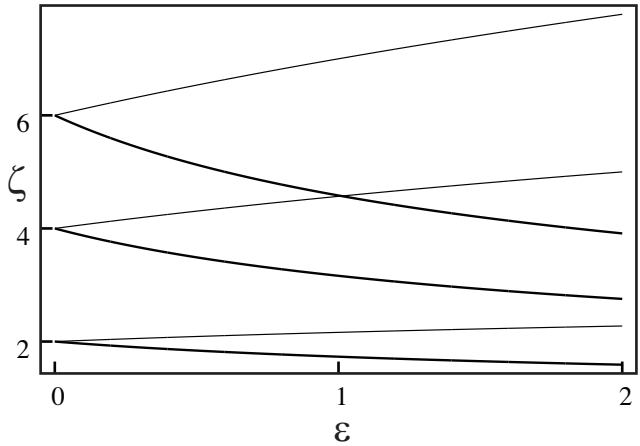}
\caption{Behavior of the exponents
$\zeta_l$ ($l=2$, 4 and 6 from the bottom to the top)
from Eq.~(\protect\ref{root-te})
{\it vs} $\varepsilon $ in three dimensions for $\alpha=0$ (thin lines)
and $\alpha=\infty$ (thick lines).}
\label{fig1}
\end{figure}

\begin{figure}
\epsffile{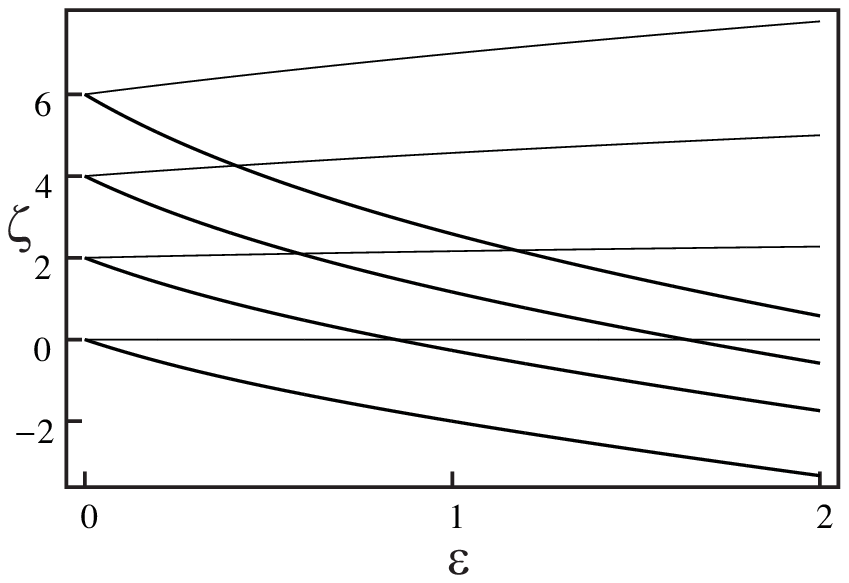}
\caption{Behavior of the exponents
$\zeta_l$ ($l=0$, 2, 4 and $6$ from the bottom to the top)
from Eq.~(\protect\ref{root-de})
{\it vs} $\varepsilon $ in three dimensions for $\alpha=0$ (thin lines)
and $\alpha=\infty$ (thick lines).}
\label{fig2}
\end{figure}

\begin{figure}
\epsffile{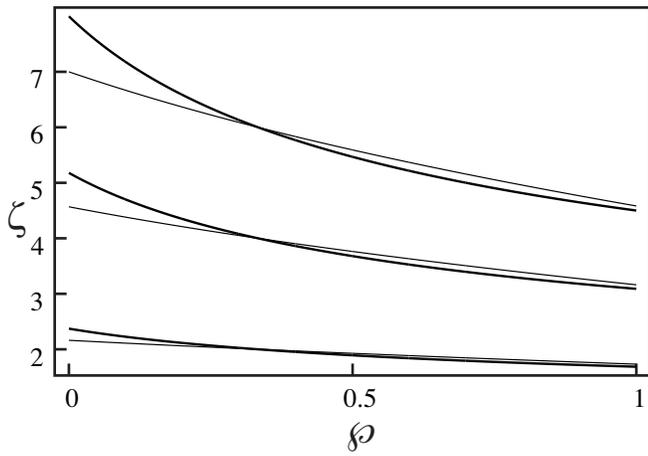}
\caption{Behavior of the exponents
$\zeta_l$ ($l=2$, 4 and 6 from the bottom to the top)
from Eq.~(\protect\ref{root-te})
{\it vs} $\wp=\alpha/(d-1+\alpha)$ for $\varepsilon =1$,
$d=3$ (thin lines) and $d=2$ (thick lines).}
\label{fig3}
\end{figure}

\end{document}